\documentclass[aps,prb,reprint,twocolumn,superscriptaddress,showpacs,floatfix,longbibliography]{revtex4-1}

\usepackage{amsmath}
\usepackage{color}
\usepackage{mathrsfs}
\usepackage{dcolumn}
\usepackage{bm}
\usepackage{multirow}
\usepackage{graphicx}
\usepackage{rotating}
\usepackage{layout}
\usepackage[version=3]{mhchem}
\usepackage{braket}
\usepackage{longtable}
\usepackage{setspace}

\allowdisplaybreaks
\usepackage{xfrac}

\usepackage[colorlinks,citecolor=blue,urlcolor=blue,bookmarks=false,hypertexnames=true]{hyperref}

\begin{document}
\title{Correlated reference-assisted variational quantum eigensolver}
\date{\today}
\author{Nhan Trong Le}
\affiliation{University of Science, Vietnam National University, Ho Chi Minh City 700000, Vietnam}
\author{Lan Nguyen Tran}
\email{tnlan@hcmiu.edu.vn}
\affiliation{Department of Physics, International University, Ho Chi Minh City 700000, Vietnam}
\affiliation{Vietnam National University, Ho Chi Minh City 700000, Vietnam}
\date{\today}

\begin{abstract}
We propose an active-space approximation to reduce the quantum resources required for variational quantum eigensolver (VQE). Starting from the double exponential unitary coupled-cluster ansatz and employing the downfolding technique, we arrive at an effective Hamiltonian for active space composed of the bare Hamiltonian and a correlated potential caused by the internal-external interaction. The correlated potential is obtained from the one-body second-order M{\o}ller-Plesset perturbation theory (OBMP2), which is derived using the canonical transformation and cumulant approximation. Considering different systems with singlet and doublet ground states, we examine the accuracy in predicting both energy and density matrix (by evaluating dipole moment). We show that our approach can dramatically outperform the active-space VQE with an uncorrelated Hartree-Fock reference. 
\end{abstract}

\maketitle

\section{Introduction}
Quantum computing is highly promising for simulating
challenging molecules and materials \cite{bauer2020quantum,mcardle2020quantum}. It is most likely beneficial for systems with strong correlations where perturbative techniques fail\cite{cao2019quantum,motta2021emerging}. Unfortunately, current quantum hardware called noisy intermediate-scale quantum (NISQ) is limited due to noise and decoherence\cite{de2021materials}. The variational quantum eigensolver (VQE) has been proposed as a low-depth quantum algorithm \cite{peruzzo2014variational,mcclean2016theory,kandala2017hardware} to take advantage of NISQ computers. It is a hybrid quantum-classical method that must be carried out on quantum and classical computers. On the quantum computer, quantum states depending on a set of variational parameters are prepared, and the expectation value of the Hamiltonian is then measured. Since most operations on a quantum computer are unitary, unitary coupled-cluster (UCC) wavefunction \cite{bartlett1989alternative,taube2006new,romero2018strategies,anand2022quantum} has been proposed as a low-circuit-depth state-preparation ansatz for VQE. Next, the set of variational variables is optimized on classical computers, and the loop is repeated until it converges. 

However, the VQE applicability is limited by the dimensionality of many-body systems associated with the number of variational variables and the circuit depths. Recently, many methodologies have been developed to reduce the resources required in VQE. Izmaylov and co-workers \cite{QCC2018,QCC2020} developed the qubit coupled cluster approach that works directly in the qubit space. They showed that QCC with factorization allows for very efficient use of quantum resources in terms of the number of coupled cluster operators. Grimsley and co-workers proposed the adaptive-VQE (ADAPT-VQE) ansatz\cite{grimsley2019adaptive} that is progressively built by subsequently incorporating into it the operators that contribute most to minimizing the VQE energy towards the ground-state energy. Later on, several groups  extended  ADAPT-VQE to make the algorithm more efficient\cite{tang2021qubit,zhang2021mutual}. While ADAPT-VQE was shown to outperform standard ansatz, its iterative nature makes calculations more costly. Freericks and co-workers, on the other hand, have devised the factorized UCC ansatz for VQE\cite{fac-UCC,xu2022decomposition} that employs a Taylor expansion in the small amplitudes, trading off circuit depth for additional measurements. Strong correlations were considered by performing the expansion about a small set of UCC factors that are treated exactly. There have been other adaptive VQE methods inspired by classical quantum chemistry methods\cite{selectiveUCC2022,projectiveVQE2021}.

Alternatively, one can reduce the dimensionality of the many-body Hamiltonian used in VQE by partitioning the whole system into smaller active spaces that can be handled by quantum computing. Kowalski and co-workers have employed the downfolding framework based on the double UCC (DUCC) to construct effective active-space Hamiltonians \cite{bauman2019downfolding,kowalski2021dimensionality} that integrate out high-energy Fermionic degrees of freedom while being capable of reproducing exact energy of quantum systems. To this end, one needs to define subsets of excitations either entirely within the subsets considered or involving some external orbitals. The approach can capture the effect of the whole orbital space in small-size active spaces\cite{metcalf2020resource,chladek2021variational}. Inspired by the divide-and-conquer technique in classical quantum chemistry, Nakagawa and co-workers proposed a method called deep VQE\cite{fujii2022deep,mizuta2021deep}. In the first step of this method, the whole system is divided into much smaller subsystems, each of which is solved independently using VQE. In the next step, the ground states of subsystems are used as a basis with reduced degrees of freedom to construct an effective Hamiltonian considering the inter-subsystem interactions. The resulting effective Hamiltonian is finally solved using VQE. 

Quantum embedding frameworks, such as density matrix embedding theory (DMET)\cite{mineh2022solving,tilly2021reduced,ralli2022scalable}, dynamical mean-field theory (DMFT)\cite{keen2020quantum,bauer2016hybrid}, and density functional embedding theory\cite{ralli2022scalable,gujarati2022quantum}, have been employed to make quantum computation feasible for real molecules and materials. Recently, Rossmannek et al. have demonstrated the performance of the VQE embedding into classical mean-field methods, including Hartree-Fock (HF) and density functional theory (DFT)\cite{rossmannek2021quantum}. Those authors restricted the quantum computation to a critical subset of molecular orbitals, whereas the remaining electrons provide the embedding potential computed using classical mean-field theories. The proposed embedding schemes obtained significant energy corrections to the HF and DFT reference for several simple molecules in their strongly correlated regime and larger systems of oxirane size. However, most of those calculations were limited to a minimal basis. It may be interesting to explore the performance of the active-space VQE approach on larger basis sets.

In the present work, we propose an active-space VQE approximation where VQE is naturally embedded in a correlated mean-field reference. To this end, we start from the double exponential UCC ansatz consisting of excitation operators from internal and external contributions. Employing the downfolding technique, we arrive at an effective Hamiltonian for active space composed of the bare Hamiltonian and a correlated potential describing the internal-external interaction. This correlated potential is derived using our recently-developed correlated mean-field theory called one-body second-order M{\o}ller-Plesset perturbation theory (OBMP2)\cite{OBMP2-JCP2013,tran2021improving,tran_PCCP_22}. Unlike standard MP2, OBMP2 is self-consistent, meaning that it can bypass challenges caused by the non-iterative nature of standard MP2. Details of the procedure are given in Section~\ref{sec:theo}. We demonstrate the performance of our approach by considering different systems with singlet and doublet ground states in Section~\ref{sec:result}. We examine the accuracy in predicting both energy and density matrix (by evaluating dipole moment). We show that the active-space VQE with the correlated reference outperforms the standard active-space VQE. 

\section{Theory and computation}{\label{sec:theo}}
\subsection{Variational quantum eigensolver: UCC ansatz}

VQE relies on the variational principle, which states that the ground-state energy $E_0$ is always less than or equal to the expectation value of Hamiltonian $\hat{H}$  calculated with the trial wavefunction $\left|\psi\right>$
\begin{align}
    E_0 \leq \frac{\left<\psi\right| \hat{H} \left|\psi\right>}{\left<\psi\right|\left|\psi\right>}
\end{align}
with the molecular Hamiltonian as
\begin{align}
  \hat{H} = \hat{h} + \hat{v} =  \sum_{pq}h^{p}_{q} \hat{a}_{p}^{q} + \tfrac{1}{2}\sum_{pqrs}g^{p r}_{q s}\hat{a}_{p r}^{q s}\label{eq:h1}
\end{align}
where $\left\{p, q, r, \ldots \right\}$ indices refer to general ($all$) spin orbitals. The excitation operators $\hat{a}_{p}^{q}$ and  $\hat{a}_{p r}^{q s}$ are defined as: $\hat{a}_{p}^{q} =  \hat{a}^{\dagger}_p \hat{a}_q$ and $\hat{a}_{pr}^{qs} =  \hat{a}^{\dagger}_p \hat{a}^{\dagger}_r \hat{a}_s \hat{a}_q$, where $\hat{a}^{\dagger}_p$ and $\hat{a}_q$ are fermionic creation and annihilation operators, respectively, 
One- and two-electron integrals, $h_{pq}$ and $v_{pq}^{rs}$, are in turn defined as
\begin{align*}
    h_{pq} = \int \phi_p^*(\vec{r_1})\left(-\tfrac{1}{2}\nabla^2-\sum_{I=1}^{M}\frac{Z_{I}}{r_{1I}}\right)\phi_q(\vec{r_1 }) d\vec{r_1 }, \\
    v_{pq}^{rs}=\int \phi_p^*(\vec{r_1 }) \phi_q^*(\vec{r_2})\frac{1}{r_{12}} \phi_s(\vec{r_2}) \phi_r(\vec{r_1}) d\vec{r_1 } d\vec{r_2},
\end{align*}
where $Z_I$ is the nuclear charge of atom $I$, and $r_{1I} = \left|\vec{r_1}-\vec{R_I}\right|$ and $r_{12} = \left|\vec{r_1}-\vec{r_2}\right|$.

The objective of the VQE is to minimize the expectation value of the Hamiltonian with respect to $\left|\psi\right>$. 
To implement this optimization problem on the quantum computer, one has to start by defining a wavefunction ansatz that can be expressed as a series of quantum gates. To this end, we express $\left|\psi\right>$ as the application of a parametrized unitary operator $U(\boldsymbol \theta)$ to an initial state $\left|\boldsymbol 0 \right>$ for $N$ qubits, with $\boldsymbol \theta$ representing a set of parameters varying values in $\left(-\pi, \pi\right]$. Given that trial wavefunctions, $\left|\psi\right>$, are necessarily normalized, we can now write the VQE optimization problem as follows.
\begin{align}
    E_{\text{VQE}} = \min_{\boldsymbol \theta} \left<\boldsymbol 0\right| U^\dagger(\boldsymbol \theta) \hat{H} U(\boldsymbol \theta) \left|\boldsymbol 0\right> \label{eq:E-vqe}
\end{align}

The unitary coupled cluster (UCC) ansatz is perhaps the most widely-used ansatz for VQE and given as
\begin{align}
    \left|\psi_\text{UCC} \right> = e^{\hat{A}} \left|\boldsymbol 0\right>,
\end{align}
where $\left|\boldsymbol 0 \right>$ is the HF reference and $\hat{A}$ is an anti-Hermitian combination of particle-hole excitation and de-excitation:
\begin{align}
    \hat{A} &= \hat{T} - \hat{T}^\dagger \label{eq:A_op}\\
    \hat{T} &= \sum_{i}^{occ}\sum_{a}^{vir} T_{i}^{a} \hat{a}_a^i + \sum_{ij}^{occ}\sum_{ab}^{vir} T_{ij}^{ab} \hat{a}_{ab}^{ij} + ... \label{eq:T_exc} 
\end{align}
where $\left\{i, j, k, \ldots \right\}$ indices refer to occupied ($occ$) spin orbitals and
$\left\{a, b, c, \ldots \right\}$ indices refer to virtual ($vir$) spin orbitals. 
The amplitudes $T_{i}^{a}$ and $T_{ij}^{ab}$ are parameterized into rotation angles $\boldsymbol\theta$ that are variationally optimized. Because the computational cost scales exponentially with the system size, the excitation operator is usually truncated at single and double excitations, resulting in UCC singles and doubles (UCCSD). Implementing UCC ansatz on quantum circuits requires the decomposition of operations into one- and two-qubit gates. However, such a decomposition produces the number of gates growing rapidly with the number of qubits. Thus, to implement $N$-qubit unitary operators, one needs to invoke an approximation such as Suzuki-Trotter  \cite{suzuki1976generalized} given by
\begin{align}
    e^{\hat{T} - \hat{T}^\dagger} \simeq \left(\prod_{i=1} e^{\frac{\theta_i}{\rho}\left(\hat{T}_i - \hat{T}_i^\dagger\right)} \right)^{\rho} \label{eq:trotter-ucc}
\end{align}
where $\theta_i$ is the amplitude weight associated with the $i^{\text{th}}$ excitation operator and $\rho$ is the Trotter number (order of decomposition).

VQE employs HF wavefunction as a reference, and orbitals are fixed during calculation. However, it is well-known that HF orbitals are not optimal for correlated methods. Recently, several works have proposed the orbital-optimized VQE (OO-VQE) method, in which orbitals are optimized by making the energy stationary with respect to orbital rotation parameters\cite{sokolov2020-ooVQE,mizukami2020-ooVQE,yalouz2021-SA-ooVQE,ratini2022-WAHTOR}. This approach requires the orbital gradient of VQE energy, demanding additional computational costs.    

\subsection{Correlated mean-field theory: OBMP2}{\label{sec:obmp2}}
Let us recap the OBMP2 theory, whose formulation details are presented in Refs.~\citenum{tran2021improving} and ~\citenum{OBMP2-JCP2013}. The OBMP2 approach was derived through the canonical transformation \cite{CT-JCP2006,CT-JCP2007,CT-ACP2007,CT-JCP2009,CT-JCP2010,CT-IRPC2010}, in which an effective Hamiltonian that includes dynamic correlation effects is achieved by a similarity transformation of the molecular Hamiltonian $\hat{H}$ using a unitary operator $e^{\hat{A}}$ :
\begin{align}
\hat{\bar{H}} = e^{\hat{A}^\dagger} \hat{H} e^{\hat{A}},
\label{Hamiltonian:ct}
\end{align}
with the anti-Hermitian excited operator $\hat{A}$ defined as in Eq~\ref{eq:A_op}. In OBMP2, the cluster operator $\hat{A}$ is modeled such that including only double excitation. 
\begin{align}
  \hat{A} = \hat{A}_\text{D} = \tfrac{1}{2} \sum_{ij}^{occ} \sum_{ab}^{vir} T_{ij}^{ab}(\hat{a}_{ab}^{ij} - \hat{a}_{ij}^{ab}) \,, \label{eq:op1}
\end{align}
with the MP2 amplitude 
\begin{align}
  T_{i j}^{a b} =  \frac{g_{i j}^{a b} } { \epsilon_{i} + \epsilon_{j} - \epsilon_{a} - \epsilon_{b} } \,, \label{eq:amp}
\end{align}
where $\epsilon_{i}$ is the orbital energy of the spin-orbital $i$. Using the Baker–Campbell–Hausdorff transformation, the OBMP2 Hamiltonian is defined as
\begin{align}
  \hat{H}_\text{OBMP2} = \hat{H}_\text{HF} + \left[\hat{H},\hat{A}_\text{D}\right]_1 + \tfrac{1}{2}\left[\left[\hat{F},\hat{A}_\text{D}\right],\hat{A}_\text{D}\right]_1.
 \label{eq:h2}
\end{align}
with
\begin{align}
  \hat{H}_\text{HF} &= \hat{F} + C = \hat{h} + \hat{v}_{\text{HF}} +C \label{eq:h1hf}
\end{align}
Where $\hat{h}$ is the one-electron Hamiltonian defined in Eq~\ref{eq:h1}. The HF potential $\hat{v}^{\text{HF}}$ and the constant $C$ is given as:
\begin{align}
    \hat{v}_{\text{HF}} &= \sum_{pq}^{all}\sum_{i}^{occ}\left(g^{p i}_{q i} - g^{p i}_{i q} \right) \hat{a}_{p}^{q} \label{eq:vhf}\\
    C &=  \,\, \sum_{ij}^{occ} \left(g^{ij}_{ji} - g_{ij}^{ij} \right) \,. 
\end{align}
In Eq.\ref{eq:h2}, commutators with the subscription 1, $[\ldots]_1$, involve one-body operators and constants that are reduced from many-body operators using the cumulant approximation\cite{cumulant-JCP1997,cumulant-PRA1998,cumulant-CPL1998,cumulant-JCP1999}. Doing some derivation, we eventually arrive at the OBMP2 Hamiltonian as follows
\begin{align}
  \hat{H}_\text{OBMP2} = & \,\, \hat{H}_\text{HF} + \hat{v}_\text{OBMP2} \label{eq:h4}
\end{align}
where $\hat{v}_\text{OBMP2}$ is a correlated potential composing of one-body operators. The working expression is given as
\begin{align}
\hat{v}_{\text{OBMP2}} = &  \overline{T}_{i j}^{a b} \left[ f_{a}^{i} \,\hat{\Omega}\left( \hat{a}_{j}^{b} \right) 
  + g_{a b}^{i p} \,\hat{\Omega} \left( \hat{a}_{j}^{p} \right) - g^{a q}_{i j} \,\hat{\Omega} \left( \hat{a}^{b}_{q} \right) \right] \nonumber \\  &- 2 \overline{T}_{i j}^{a b}g^{i j}_{a b} 
   + \,f_{a}^{i}\overline{T}_{i j}^{a b}\overline{T}_{j k}^{b c} \,\hat{\Omega} \left(\hat{a}_{c}^{k} \right) \nonumber \\ 
     &+  f_{c}^{a}T_{i j}^{a b}\overline{T}_{i l}^{c b} \,\hat{\Omega} \left(\hat{a}^{l}_{j} \right) + f_{c}^{a}T_{i j}^{a b}\overline{T}_{k j}^{c b} \,\hat{\Omega} \left(\hat{a}^{k}_{i} \right) \nonumber \\ 
     &-  f^{k}_{i}T_{i j}^{a b}\overline{T}_{k l}^{a b} \,\hat{\Omega} \left(\hat{a}_{l}^{j} \right)
     -  f^{p}_{i}T_{i j}^{a b}\overline{T}_{k j}^{a b} \,\hat{\Omega} \left(\hat{a}^{p}_{k} \right) \nonumber \\ 
     & +  f^{k}_{i} T_{i j}^{a b}\overline{T}_{k j}^{a d} \,\hat{\Omega}\left(\hat{a}_{b}^{d} \right) +  f_{k}^{i}T_{i j}^{a b}\overline{T}_{k j}^{c b} \,\hat{\Omega} \left(\hat{a}_{a}^{c} \right) \nonumber \\ 
     &-  f_{c}^{a}T_{i j}^{a b}\overline{T}_{i j}^{c d} \,\hat{\Omega} \left(\hat{a}^{b}_{d} \right) \,
     - f_{p}^{a}T_{i j}^{a b}\overline{T}_{i j}^{c b} \,\hat{\Omega} \left(\hat{a}^{p}_{c} \right) \nonumber \\
     & - 2f_{a}^{c}{T}_{i j}^{a b}\overline{T}_{i j}^{c b} +  2f_{i}^{k}{T}_{i j}^{a b}\overline{T}_{k j}^{a b}. \label{eq:vobmp2} 
\end{align}
with $\overline{T}_{ij}^{ab} = {T}_{ij}^{ab} - {T}_{ji}^{ab}$, the symmetrization operator $\hat{\Omega} \left( \hat{a}^{p}_{q} \right) = \hat{a}^{p}_{q}  + \hat{a}^{q}_{p}$, and the Fock matrix 
\begin{align}
    f_p^q = h_p^q + \sum_{i}^{occ}\left(g^{p i}_{q i} - g^{p i}_{i q} \right).
\end{align}
Note that, for convenience, we have used Einstein's convention in Eq.~\ref{eq:vobmp2} to present the summations over repeated indices. We rewrite $\hat{H}_\text{OBMP2}$ (Eqs. \ref{eq:h2} and \ref{eq:h4}) in a similar form to Eq. \ref{eq:h1hf} for $\hat{H}_\text{HF}$ as follows:
\begin{align}
  \hat{H}_\text{OBMP2} = & \hat{\bar{F}} + \bar{C} \label{eq:h5}
\end{align}
with $\hat{\bar{F}} =  \bar{f}^{p}_{q} \hat{a}_{p}^{q}$.
$\bar{f}^{p}_{q}$ is so-called {\it correlated} Fock matrix and written as
\begin{align}
\bar{f}^{p}_{q} &= f^{p}_{q} + v^{p}_{q}. \label{eq:corr-fock}
\end{align}
$v^{p}_{q}$ is the matrix representation of the one-body operator $\hat{v}_{\text{OBMP2}}$, serving as the correlation potential altering the uncorrelated HF picture. We update the MO coefficients and energies by diagonalizing the matrix $\bar{f}^{p}_{q}$, leading to orbital relaxation in the presence of dynamic correlation effects. The formal scaling of OBMP2 is similar to standard MP2 ($N^5$). The OBMP2 method is implemented within a local version of PySCF\cite{pyscf-2018}. 

\subsection{VQE with OBMP2 reference: downfolding approach}

We can see that both UCC and OBMP2 are formulated using a unitary exponential operator $e^{\hat{A}}$ (Eqs~\ref{eq:A_op} and \ref{eq:op1}), implying that one can combine these two naturally. Partitioning the whole orbital space into active and inactive spaces, one can write the double unitary CC (DCC) ansatz as\cite{bauman2019downfolding}
\begin{align}
    \left| \boldsymbol \phi \right> =  e^{\hat{A}_{\text{ext}}} e^{\hat{A}_{\text{int}}}  \left| \boldsymbol 0 \right>
\end{align}
Here, the internal (int) defines excitations within the active space, and the external (ext) is the remaining excitations involving at least one inactive space orbital.  

The total energy can be written as:
\begin{align}
    E &=  \left< \boldsymbol 0 \right| e^{\hat{A}_{\text{int}}^\dagger} e^{\hat{A}_{\text{ext}}^\dagger} \hat{H} e^{\hat{A}_{\text{ext}}} e^{\hat{A}_{\text{int}}}  \left| \boldsymbol 0 \right> \nonumber \\
    &= \left< \boldsymbol 0 \right| e^{\hat{A}_{\text{int}}^\dagger} \hat{\bar{H}}_{\text{ext}} e^{\hat{A}_{\text{int}}}  \left| \boldsymbol 0 \right> 
\end{align}
where $\hat{\bar{H}}_{\text{ext}}$ is the effective Hamiltonian and defined as,
\begin{align}
 \hat{\bar{H}}_{\text{ext}} &= e^{\hat{A}_{\text{ext}}^\dagger} \hat{H} e^{\hat{A}_{\text{ext}}} \nonumber \\
 &\simeq \hat{H} + \left[\hat{H},\hat{A}_\text{ext}\right] + \tfrac{1}{2}\left[\left[\hat{H},\hat{A}_\text{ext}\right],\hat{A}_\text{ext}\right].\label{eq:hext}
\end{align}
Here, we have truncated the BCH expansion at the second order. Letting $\hat{A}_{\text{ext}} = \hat{A}_{\text{D}}^{\text{ext}}$ that includes only double excitations with at least one inactive index  and employing the OBMP2 approximation for the last two terms in Eq~\ref{eq:hext}, we arrive at  %
\begin{align}
 \hat{\bar{H}}_{\text{ext}} &= \hat{H} + \hat{v}_{\text{OBMP2}}^{\text{ext}}, \label{eq:hext-obmp2}
\end{align}
where the amplitudes in $\hat{v}_{\text{OBMP2}}^{\text{ext}}$ carry at least one active index for the amplitudes. We further use the active space approximation, in which the effective Hamiltonian \ref{eq:hext-obmp2} is truncated within active space and defined by strings of creation/annihilation operators carrying only active spin orbitals. One can rewrite Eq.~\ref{eq:hext-obmp2} as
\begin{align}
\hat{\bar{H}}^{\text{act}}_{\text{ext}} = \sum_{pq \in \text{act}}\bar{h}^{p}_{q} \hat{a}_{p}^{q} + \tfrac{1}{2}\sum_{pqrs \in \text{act}}g^{p r}_{q s}\hat{a}_{p r}^{q s}
\label{eq:heff}
\end{align}
Here we define $\bar{h}_{pq} = h_{pq} + v_{pq}^{\text{ext}}$ as the effective one-electron integral including the uncorrelated one-electron part $h_{pq}$ and the OBMP2 part $v_{pq}^{\text{ext}}$ (the matrix representation of $\hat{v}_{\text{OBMP2}}^{\text{ext}}$). 
We emphasize that by carrying at least one inactive orbital, the amplitudes in $v_{pq}^{\text{ext}}$ capture the dynamical correlation outside the active space. One of the main advantages of using OBMP2 downfolding is that the effective correlated potential includes only one-body operators that can be involved into the one-body term of effective Hamiltonian. In such a way, while  coefficients of one-body term are different from those in bare Hamiltonian, there is no additional cost when implementing effective Hamiltonian on quantum computers. 

The total density matrix is the sum of the inactive core and active space as follows
\begin{align}
    D_{\text{total}} = D_{\text{inact}} + D_{\text{act}}. 
\end{align}
The active density matrix is obtained from VQE wave-functions, whereas the inactive part is idempotent and evaluated using either HF or OBMP2 molecular orbitals. We have shown that the difference between OBMP2 density matrices evaluated using relaxed molecular orbitals (HF-like) and double excitation amplitudes (MP2-like) is insignificant in many cases \cite{tran2021improving}.

Calculations start by running OBMP2 for the whole system and selecting active space using OBMP2 orbitals. We would stress that the scaling of OBMP2 method is similar to that of standard MP2, which is $N^5$. VQE is then used for the active space with the effective Hamiltonian \ref{eq:heff}. As recently discussed by Bauman and Kowalski \cite{bauman2022coupled}, the accuracy of ground-state energies from downfolding methods strongly depends on various approximations. First, the effective Hamiltonian is truncated at the second-order BCH expansion (Eq.~\ref{eq:hext}), implying that the spectrum of the original Hamiltonian can be only approximately obtained. Second, if amplitudes of $\hat{A}_{\text{ext}}$ are large, VQE in active space may not adequately capture the static correlation. Because VQE/OBMP2 method can be considered the perturb-then-diagonalize approach, the divergence of the perturbative step may lead to the failure of calculation. As we show later, the self-consistency of OBMP2 can partially remove the large-value issue of MP2 amplitudes.   

The classical calculation is carried out using PySCF\cite{pyscf-2018}, and the quantum part is done using the Qiskit package\cite{Qiskit}. The operators are transformed to qubit space using the Jordan-Wigner mapping, and the ansatz is approximated with a single Trotter step ($\rho = 1$ in Eq.~\ref{eq:trotter-ucc}). MP2 amplitudes are used as initial guesses of UCC parameters. We employed the noiseless Aer simulator together with the SLSQP optimizer. The threshold for the convergence of VQE energy is $10^{-12}$ a.u.

\section{Results and discussion}{\label{sec:result}}

\subsection{Full-space VQE with OBMP2 orbitals}

\begin{figure*}[t!]
  \includegraphics[width=16cm,]{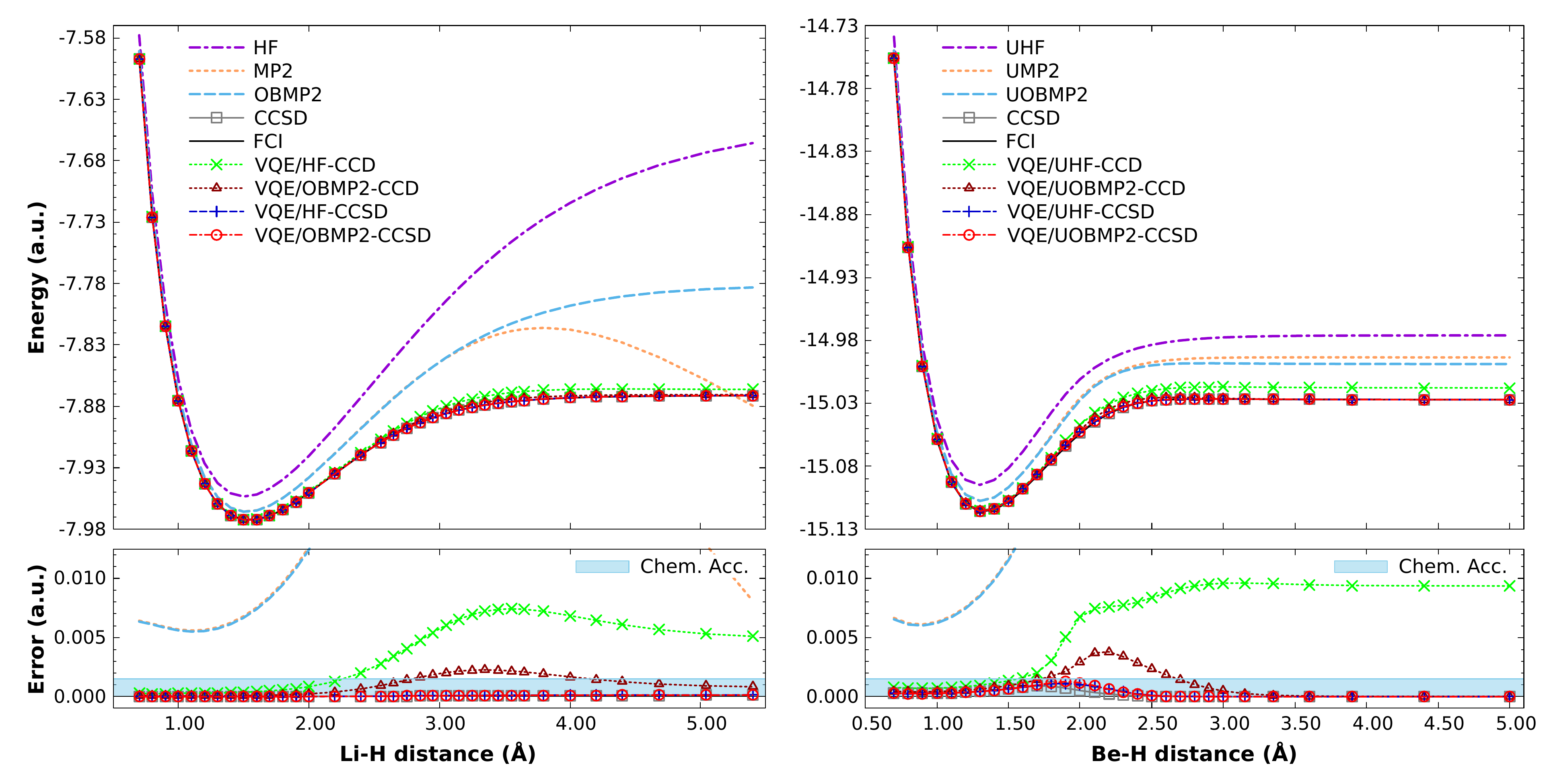}
  \caption{\normalsize Potential energy curves of LiH (left) and BeH (right) in STO-6G. VQE with different ansatz (UCCD and UCCSD) was performed for the full orbital space optimized using HF and OBMP2. The chemical accuracy ($\leq 1.5$ mHa) is represented by the blue region.}
  \label{fig:lih-beh-sto}
\end{figure*}

Several authors have shown that orbital relaxation is important to reduce VQE errors \cite{sokolov2020-ooVQE,mizukami2020-ooVQE,yalouz2021-SA-ooVQE,ratini2022-WAHTOR} and the number of qubits \cite{OptOrbVQE2023}. In those studies, the energy of VQE is minimized concerning both cluster amplitudes and orbitals, resulting in a self-consistency that demands higher computational costs than standard VQE. It is thus interesting to examine whether correlated orbital reference pre-optimized using a lower-level method can improve the accuracy of ``single-shot'' VQE. Here, we performed VQE only once on OBMP2 correlated orbitals. 

\begin{figure}[h!]
  \includegraphics[width=8cm,]{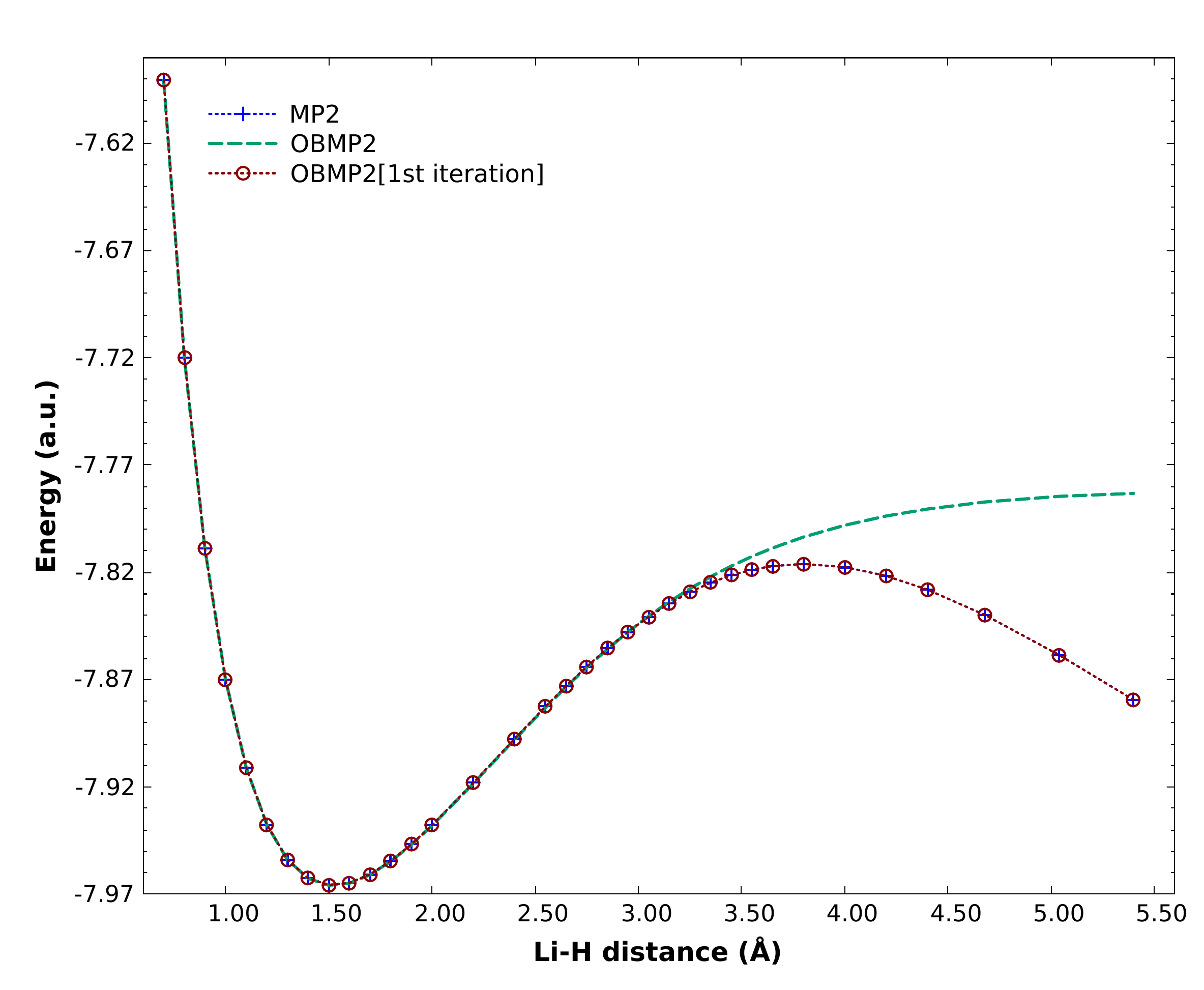}
  \caption{\normalsize Potential energy curves of LiH in STO-6G obtained from OBMP2, the first iteration of OBMP2, and MP2.}
  \label{fig:lih-sto}
\end{figure}

In Figure~\ref{fig:lih-beh-sto}, we plot the potential energy curves of LiH (left panel) and BeH (right panel) in the STO-6G basis. Note that we have used unrestricted methods for BeH that has the doublet ground state. In general, classical CCSD is nearly exact for these small systems. MP2 and OBMP2 energies are almost identical around equilibrium geometries and close to the FCI reference. Surprisingly, for LiH in the stretching regime, while restricted MP2 breaks down, restricted OBMP2 yield a curve nearly parallel to the FCI reference. This is thanks to the self-consistency of OBMP2 \cite{tran2021improving}. In Figure~\ref{fig:lih-sto}, we plot potential energy curves of LiH in STO-6G obtained from OBMP2, the fitst iteration of OBMP2, and MP2. The first iteration OBMP2 and MP2 are nearly identical each other and break down at the stretching limit. After the self-consistency, OBMP2 can bypass the failure of MP2 and describe the dissociation properly.  

Regarding VQE calculations, UCCD and UCCSD ansatzs are employed with HF and OBMP2 orbtials. For both molecules, VQE-UCCSD with different orbital sets yields curves almost identical and is nearly exact with errors within the region of chemical accuracy. The difference between results obtained from different orbital sets is prominent for VQE-UCCD, particularly in the stretched regime when a strong correlation is present. 
VQE-UCCD with the OBMP2 orbitals gives errors relative to the reference several times smaller than that with the HF orbitals. 
Also, while VQE-UCCD dramatically deviates from VQE-UCCSD for HF orbitals, the deviation between these two are small for OBMP2 orbitals. 
Noticeably, VQE-CCD with OBMP2 orbitals can attain the chemical accuracy for most distances except for the region around stretching limit. In general, using OBMP2 orbitals can help to reduce errors without additional costs. Therefore, one can use VQE/OBMP2 as an approximation to the orbital-optimized VQE approach \cite{sokolov2020-ooVQE,mizukami2020-ooVQE,yalouz2021-SA-ooVQE,ratini2022-WAHTOR}.   

\subsection{Active-space VQE for sinlget molecules}

\begin{figure*}[t!]
  \includegraphics[width=16cm,]{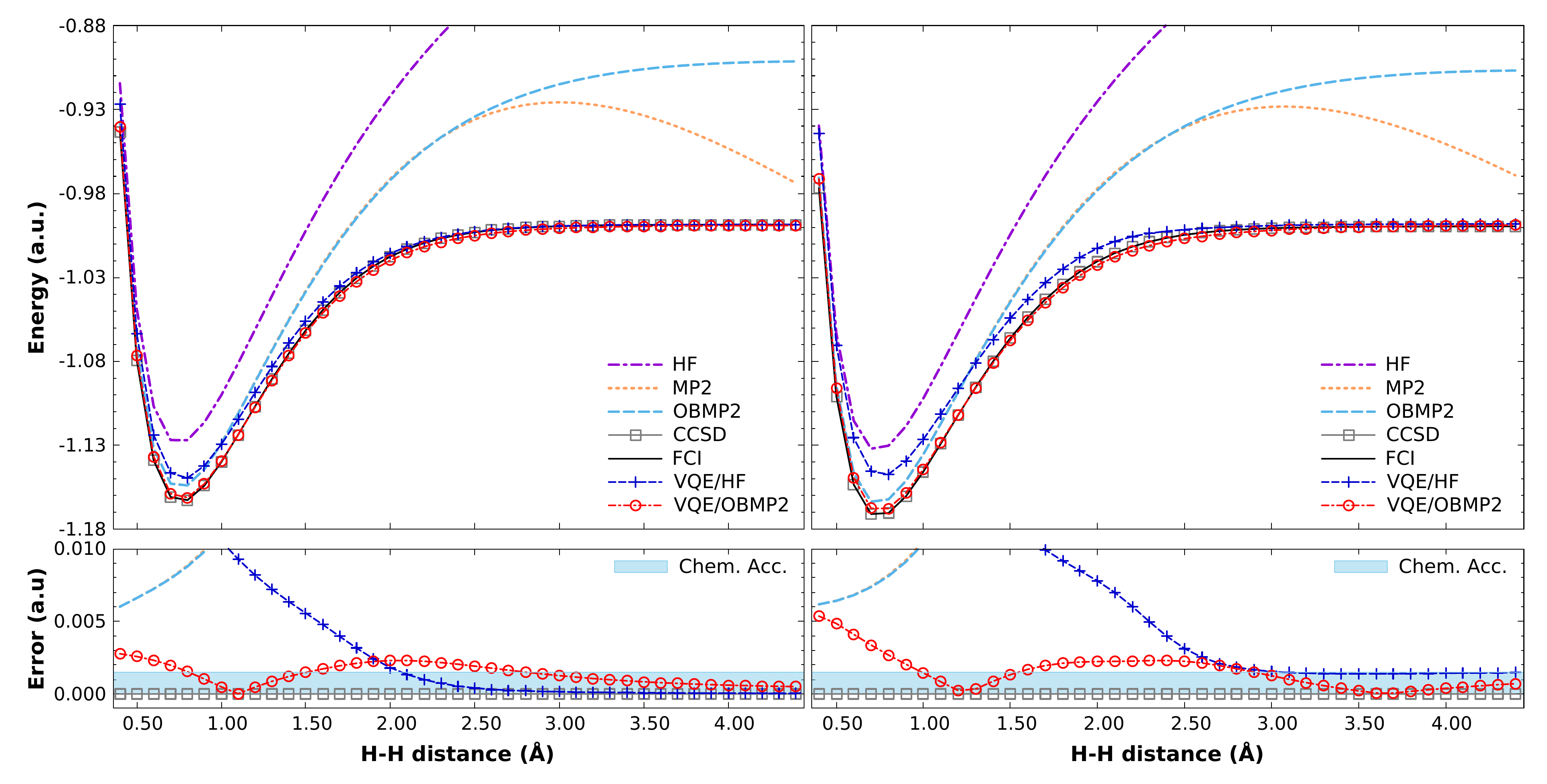}
  \caption{\normalsize Potential energy curves of H$_2$ in cc-pVDZ (left) and cc-pVTZ (right). VQE was performed in the active space of four orbitals (4o). The chemical accuracy ($\leq 1.5$ mHa) is represented by the blue region.}
  \label{fig:h4c-dz}
\end{figure*}

\begin{figure*}[t!]
  \includegraphics[width=16cm,]{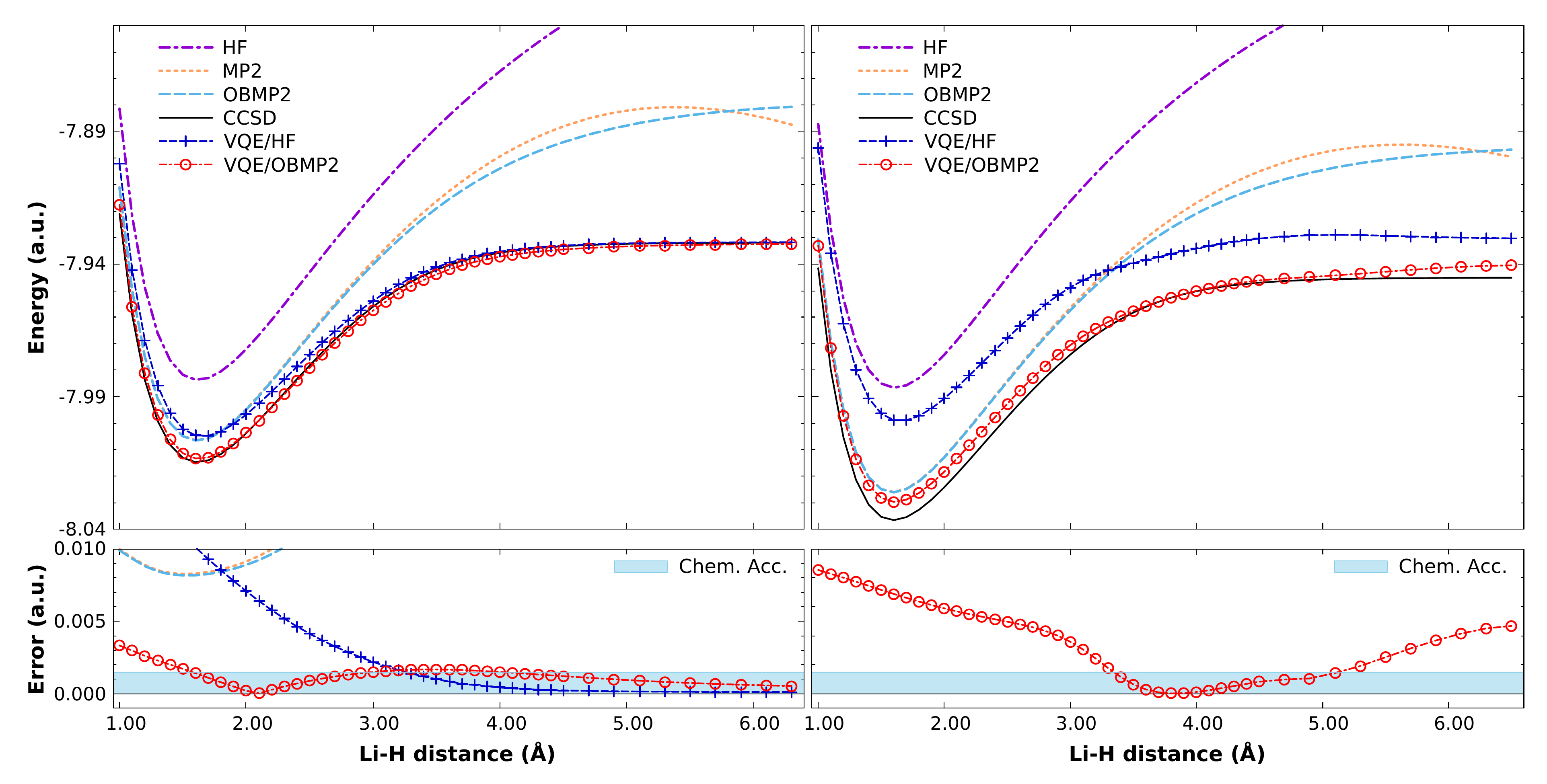}
  \caption{\normalsize Potential energy curves of LiH in cc-pVDZ (left) and cc-pVTZ (right). VQE was performed in the active space of eight orbitals (8o). For cc-pVTZ, VQE/HF errors are out the scale ($> 10$ mHa). The chemical accuracy ($\leq 1.5$ mHa) is represented by the blue region.}
  \label{fig:lih-dz}
\end{figure*}

Hereafter, unless otherwise noted, the UCCSD ansatz is used for all VQE calculations. 
Figure~\ref{fig:h4c-dz} represents potential energy curves of H$_2$ in cc-pVDZ and cc-pVTZ. VQE was performed within an active space of four orbitals (4o). For comparison, we also plot HF, MP2, OBMP2, and CCSD curves. All the methods perform similarly in the two basis sets. It is not surprising that CCSD is exact for H$_2$. While OBMP2 yields result pretty close to FCI around the equilibrium bond length, its errors are significant for stretched geometries. MP2 and OBMP2 are almost identical for short distances. However, while no divergence is observed for OBMP2, MP2 breaks down at stretched distances. Therefore, using MP2 as a low-level method in the downfolding approach may lead to suspicious results when strong correlations are present.  

Due to the lack of dynamic correlation outside the active space, VQE/HF is far from the FCI reference around the equilibrium. Its errors decrease at long distances when the strong correlation becomes dominant. On the other hand, capturing both dynamical and static correlations, VQE/OBMP2 dramatically outperforms VQE/HF and yields small errors relative to FCI for the whole range of distances considered here. Thus, the non-parallelity error (NPE), defined as the difference between the minimum and maximum errors, is smaller for VQE/OBMP2 than for VQE/HF.  

The next system we consider is LiH in cc-pVDZ and cc-pVTZ. Here, the core orbital Li $1s$ is not included in the active space and treated at the (correlated) mean-field level. VQE was performed in an active space of eight orbitals (8o) consisting of only $\sigma-$type orbitals. We employ classical restricted CCSD as the reference for this molecule. The UCCSD ansatz is used for VQE. All results are summarized in Figure~\ref{fig:lih-dz}. For the cc-pVDZ basis, while VQE/HF errors are significant around the equilibrium geometry, those are small at long distances. On the other hand, VQE/OBMP2 errors are small for the whole range of distances considered here and can attain the chemical accuracy at most distances. Dynamic correlation effects become much more critical when the basis set is enlarged to cc-pVTZ. Thus, the VQE/HF curve notoriously deviates from the CCSD reference for the whole curve. In contrast, thanks to the OBMP2 correlated potential in the effective simulated Hamiltonian, VQE errors are reduced by several orders. However, the VQE/OBMP2 calculation for cc-pVTZ still cannot reach the chemical accuracy for many distances. One needs to enlarge the active space to further reduce errors.

\begin{figure}[t!]
  \includegraphics[width=8cm]{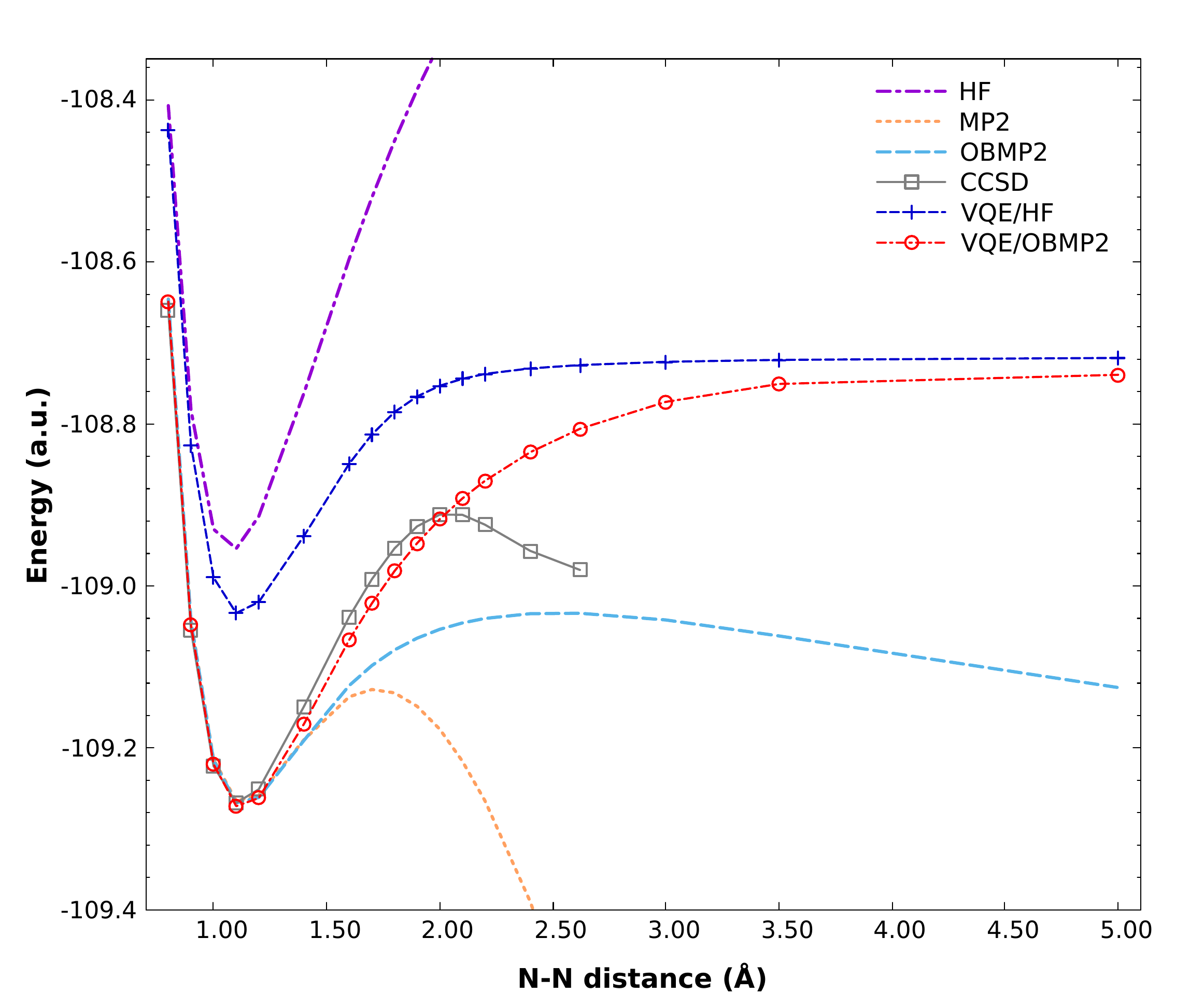}
  \caption{\normalsize Potential energy curves of N$_2$. The basis set is cc-pVDZ. VQE was performed in the active space of eight orbitals (8o).}
  \label{fig:n2}
\end{figure}

Let us now consider a more challenging molecule N$_2$. The strong correlation present in stretching the triple bond of N$_2$ makes it difficult for single-reference methods. In Figure~\ref{fig:n2}, we plot N$_2$ potential energy curves from different methods including HF, MP2, OBMP2, CCSD, VQE/HF, VQE/OBMP2. The active space for VQE calculations is eight valence orbitals (8o) composed of N 2$s$2$p$. Restricted MP2 and CCSD fail to describe the dissociation due to the strong correlation. Interestingly, restricted OBMP2 does not immediately break down at the stretching limit. However, it is still not sufficient to describe the dissociation. In general, both VQE/HF and VQE/OBMP2 describe the dissociation properly. Thanks to dynamical correlation outside the active space, VQE/OBMP2 yields a curve closer to CCSD around equilibrium than VQE/HF. In general, one can state that VQE/OBMP2 combines the advantages of both VQE and OBMP2 methods.  

\subsection{Active-space VQE for doublet molecules}

\begin{figure*}[t!]
  \includegraphics[width=16cm]{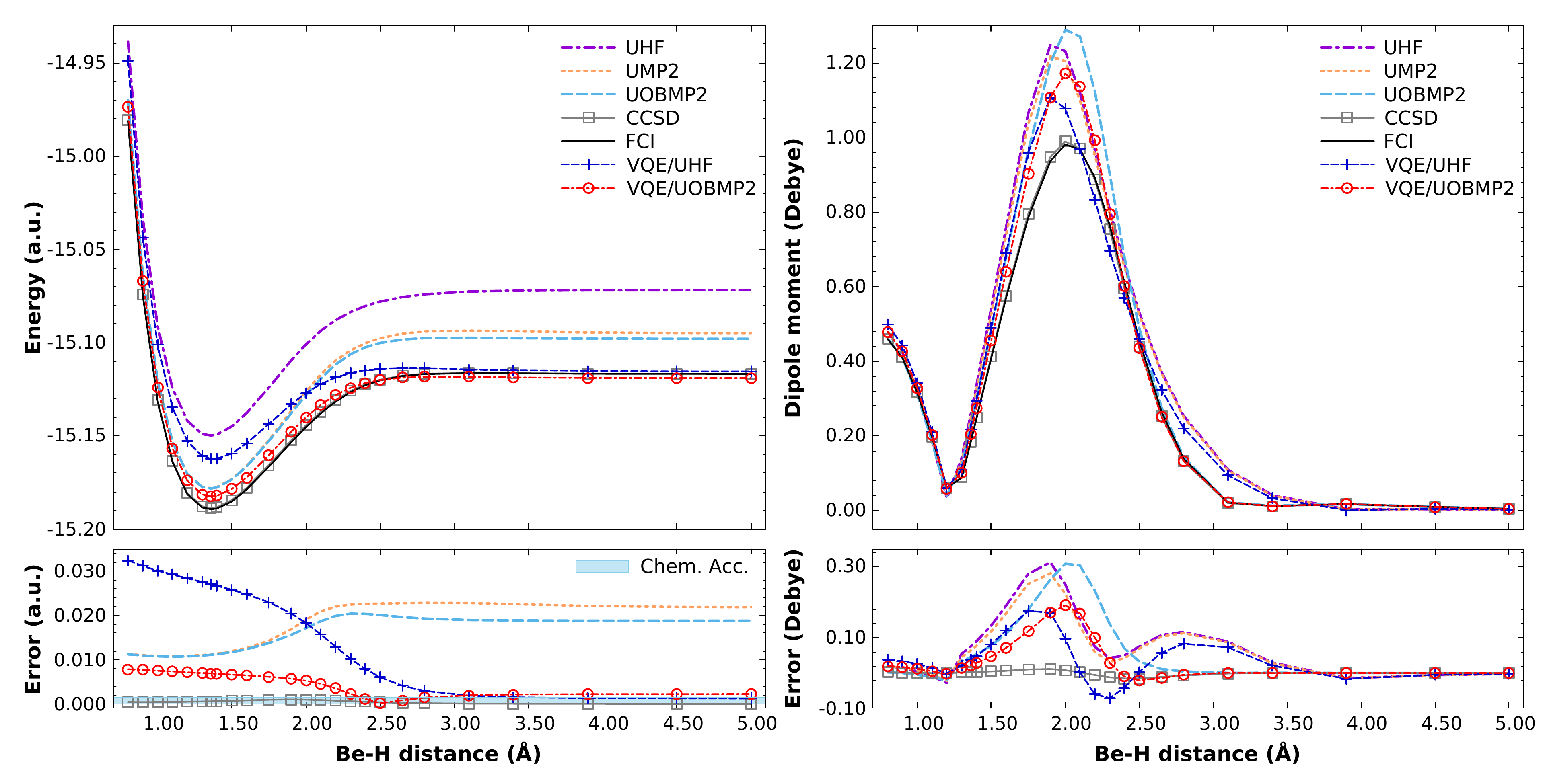}
  \caption{\normalsize Left: potential energy curves of BeH with the ground state doublet. Right: The change of BeH dipole moment. The basis set is cc-pVDZ. VQE was performed in the active space of nine orbitals (9o). The chemical accuracy ($< 1.5$ mHa) is represented by the blue region.}
  \label{fig:beh-pes-dip}
\end{figure*}

\begin{figure*}[t!]
	\includegraphics[width=16cm]{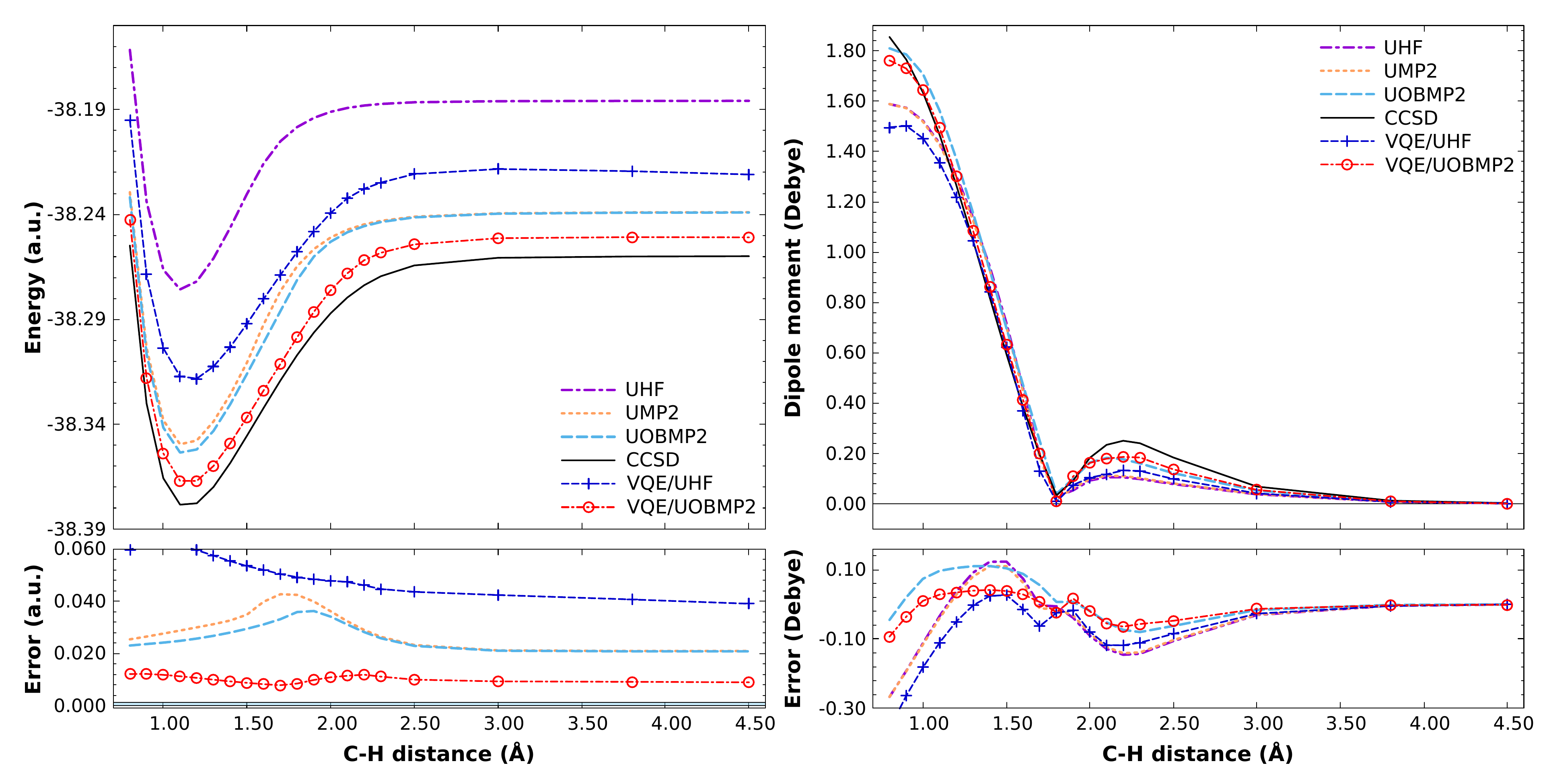}
	\caption{\normalsize Left: potential energy curves of CH with the ground state doublet. Right: The change of CH dipole moment. The basis set is cc-pVDZ. VQE was performed in the active space of nine orbitals (9o). The chemical accuracy ($\leq 1.5$ mHa) is represented by the blue region.}
	\label{fig:ch-pes-dip}
\end{figure*}

This subsection considers two systems with the ground state doublet (e.g., having one unpaired electron): BeH and CH. The unrestricted HF and OBMP2 are used as the reference for VQE. In addition to potential energy curves, we also calculate dipole moments, $\vec{\mu} = \int \vec{r}\rho(\vec{r}) d\vec{r}$, that directly measure the density matrix. We note that, in the current work, VQE density is not relaxed. The cc-pVDZ basis set is used for all calculations. 

In Figure~\ref{fig:beh-pes-dip}, we plot the potential energy curves and dipole moments of BeH in cc-pVDZ from different methods. The errors relative to the FCI reference are also presented. VQE is performed in the active space of nine orbitals (9o) including Be $2s2p$ and H $1s$. Unrestricted HF and OBMP2 describe the dissociation quite correctly with small NPEs. VQE/HF can significantly reduce HF errors at long distances, but it is still far from FCI around the equilibrium, implying the importance of dynamical correlation. VQE/OBMP2, which can capture dynamical and static correlations, can yield the potential energy curves close to FCI. Both VQE/HF and VQE/OBMP2 can attain the chemical accuracy at long distances. For short distances, one may need to enlarge the active space in VQE/OBMP2 to capture more dynamical correlation to reach the chemical accuracy. As for dipole moments, it is clear that OBMP2 yields more accurate results than HF, particularly at long distances ($R > 2.5 \r{A}$), reflecting the importance of orbital optimization in the presence of correlation for density-related properties, as we have shown recently\cite{tran2021improving,tran_PCCP_22}. Consequently, VQE with OBMP2 describes dipole moments better than that with HF.

Figure~\ref{fig:ch-pes-dip} represents the potential energy curves and dipole moments of CH from different methods and their errors relative to the CCSD reference. VQE is performed in the active space of nine orbitals (9o), composed of C $2s2p$ and H $1s$. Although unrestricted HF and OBMP2 can describe the dissociation adequately, a large NPE is observed due to a bump at the unrestricted point 1.5\r{A}. VQE/HF can reduce errors and yield the curve parallel to the FCI reference. When VQE is performed with the UOBMP2 reference (VQE/OBMP2), the errors in energy dramatically decrease with a small NPE. However, VQE/UOBMP2 with 9o is still not sufficient to reach the chemical accuracy for the whole curve. One may need to enlarge the active space further. Working on this issue is in progress. We plot the change of dipole moments when stretching the C--H bond in the right panel of Figure~\ref{fig:ch-pes-dip}. All the methods yield curves that behave similarly to the CCSD reference. Overall, VQE/OBMP2 predicts the dipole moment closest to the CCSD reference for the range of distances considered here, indicating the importance of static and dynamic correlations in accurately predicting density-related properties.

\section{Conclusion}
We have proposed an active-space approximation in which VQE is naturally embedded in the correlated mean-field reference OBMP2 derived from the downfolding technique. We partition the whole orbital space into active and inactive spaces and exploit the double exponential UCC ansatz as the product of internal and external contributions. The effective Hamiltonian for the active space is a sum of the bare Hamiltonian in the active space and a potential describing the internal-external interaction derived from OBMP2, a correlated mean-field theory recently developed by us. Considering different systems with singlet and doublet ground states in the minimal and larger basis sets, we demonstrated the accuracy of our approach in predicting energies and dipole moments. We show that the VQE with the OBMP2 reference significantly improves upon the standard active-space VQE with the uncorrelated HF reference. 

Our approach is helpful in studying realistic chemistry and materials on quantum computers. It is generally applicable to different types of UCC ansatz, such as generalized UCC\cite{lee2018generalized}, paired UCC\cite{lee2018generalized,stein2014seniority}, and pair-natural orbital-UCC\cite{kottmann2021reducing}. One can classify our approach as a perturb-then-diagonalize method. If OBMP2 cannot describe systems well, the VQE/OBMP2 method may fail. Another limitation of our method is that we are using cumulant approximation to arrive at the one-body correlated potential in OBMP2, causing the missing some dynamical correlation. Further work is to develop more sophisticated schemes of active-space selection to treat systems with large active spaces. For example, one can split the active space into smaller subspaces and treat them independently using VQE as in quantum embedding methods \cite{welborn2016bootstrap,wouters2016practical,seet-jctc2016,seet-jpcl2017}. In the current work, orbitals are only optimized at the OBMP2 level, and VQE is performed as a "single-shot" calculation. Thus, we also plan to implement the orbital relaxation in the presence of VQE correlation energy.


\bibliography{main}

\end{document}